\documentclass[review]{elsarticle}
\usepackage[T1]{fontenc}
\usepackage[utf8x]{inputenc}
\usepackage{graphics} 
\usepackage[english]{babel}
\usepackage[colorinlistoftodos]{todonotes}
\usepackage{epsfig} 
\usepackage{times} 
\usepackage{amsmath} 
\usepackage{amssymb}  
\usepackage[normalem]{ulem}
\renewcommand{\thepage}{}
\usepackage{xcolor}
\usepackage{verbatim}
\usepackage{import}
\usepackage{xcolor}
\usepackage{amssymb}
\usepackage{algorithm}
\usepackage{multirow}
\usepackage{adjustbox}
\usepackage{algpseudocode,float}
\makeatletter
\algrenewcommand\ALG@beginalgorithmic{\footnotesize}
\makeatother
\usepackage{pgfplots}
\usepackage{smartdiagram}
\usepackage{dot2texi}
\usepackage{color, colortbl}
\usepackage{mathrsfs}
\usepackage{pgfkeys}
\usepackage{diagbox}
\usepackage{mathtools}
\usepackage{url}
\usepackage{xcolor}
\usepackage{fancyhdr}
\usepackage{import}
\usepackage{tabularx,ragged2e,booktabs}
\newcolumntype{L}{>{\RaggedRight\arraybackslash}X} 
\newcolumntype{C}[1]{>{\centering\let\newline\\\arraybackslash\hspace{0pt}}m{#1}}
\makeatletter
\newcommand*{\rom}[1]{\expandafter\@slowromancap\romannumeral #1@}
\makeatother
\usepackage[encapsulated]{CJK}
\usetikzlibrary{arrows}
\usepackage{url}

\usepackage{breakurl}

\pgfplotsset{compat=1.8}
 \DeclareGraphicsExtensions{.png,.pdf}
\usepackage{array,multirow,makecell}
\usepackage{lineno}
\modulolinenumbers[5]
\begin{document}
\sloppy
\begin{frontmatter}

\title{On Byzantine Fault Tolerance in Multi-Master Kubernertes Clusters}

\author[myfirstaddress]{Gor Mack Diouf}
\author[myfirstaddress]{Halima Elbiaze\corref{mycorrespondingauthor}}
\cortext[mycorrespondingauthor]{Corresponding author}
\ead{elbiaze.halima@uqam.ca}

\address[myfirstaddress]{Universit\'e du Qu\'ebec \`A Montr\'eal, Montreal, Quebec, Canada}
\author[mysecondaryaddress]{Wael Jaafar}
\address[mysecondaryaddress]{Carleton University, Ottawa, Ontario, Canada}

\fntext[myfirstaddress]{Département d'informatique, C.P. 8888, Succursale Centre-ville 
Montréal (Québec) H3C 3P8 Canada}

\begin{abstract}
Docker container virtualization technology is being widely adopted in cloud computing environments because of its lightweight and efficiency. However, it requires adequate control and management via an orchestrator. As a result, cloud providers are adopting the open-access Kubernetes platform as the standard orchestrator of containerized applications.
To ensure applications' availability in Kubernetes, the latter uses Raft protocol's replication mechanism. Despite its simplicity, Raft assumes that machines fail only when shutdown. This failure event is rarely the only reason for a machine's malfunction. Indeed, software errors or malicious attacks can cause machines to exhibit Byzantine (i.e. random) behavior and thereby corrupt the accuracy and availability of the replication protocol. 
In this paper, we propose a Kubernetes multi-Master Robust (KmMR) platform to overcome this limitation. KmMR is based on the adaptation and integration of the BFT-SMaRt fault-tolerant replication protocol into Kubernetes environment. Unlike Raft protocol, BFT-SMaRt is resistant to both Byzantine and non-Byzantine faults.
Experimental results show that KmMR is able to guarantee the continuity of services, even when the total number of tolerated faults is exceeded. In addition, KmMR provides on average a consensus time 1000 times shorter than that achieved by the conventional platform (with Raft), in such condition. Finally, we show that KmMR generates a small additional cost in terms of resource consumption compared to the conventional platform.
\end{abstract}


\end{frontmatter}

\section{Introduction}
Faced with the continuous increase in capital expenditure 
and operating expenditure 
costs of fully reliable and available Information Technology (IT) systems, companies tend towards outsourcing their IT services to specialized companies such as cloud service providers. 
The main advantage of this strategy is to claim an excellent service quality while paying only for the necessary and consumed resources. As for the 
service provider, its purpose is to meet the needs of clients by providing the required resources when demanded. A common approach 
is to pool (or slice) its resources to share them between several clients. In this context, many challenges emerge to provide a reliable cloud environment, e.g., quality-of-service 
guarantee, resources management, and service continuity.

\textcolor{black}{In order to exploit efficiently the service provider's resources, the virtualization technology has been introduced \cite{Virtu,NFV}. The latter allows the services to see the resources, e.g., servers, routers, communication links, and data storage, in a manner that is independent from the physical infrastructure/equipment, and to use these resources based on service requirements, rather than on physical granularity. In particular, servers virtualization using containers, called also containerization, 
}
has gained popularity among cloud service providers, since it \textcolor{black}{addresses} 
issues, such as the inefficient use of resources  \cite{bernstein2014containers,peinl2016docker}. \textcolor{black}{Unlike full-hardware virtualization, such as VMware \cite{Virtu}, containerization leverages virtualization at the operating system level, hence generating a lighter overhead. 
In such system, the resource allocation unit is the container. The latter is defined as the virtual runtime environment running atop a single operating system kernel and emulating an operating system. Several implementation platforms are available for containerization, such as LXC, OpenVZ and Docker \cite{bernstein2014containers,peinl2016docker,Rizki}. Nevertheless, Docker stands out as the most interesting container-based virtualization platform as it provides the simplest lightweight and scalable way of creating and deploying containers, besides its large spectrum of use cases, including hybrid clouds \cite{HybridC}, microservices \cite{microserv}, infrastructure optimization \cite{Garg} and big data \cite{Hadoop}}.  

\textcolor{black}{In a container-based server platform, containerized applications need to be managed, i.e., a container hosting an application is dynamically deployed, run, then removed. The management of these operations in a container-based virtualization platform is called Containers Orchestration. Containers Orchestration is a complex task that requires a very light but efficient mechanism for automated deployment, scaling, and management of containers. For instance, 
} to \textcolor{black}{efficiently} manage Docker containers, cloud service providers such as Google, Docker, Mesosphere, Microsoft, VMware, IBM and Oracle 
adopted Kubernetes as their standard platform to orchestrate containerized applications \cite{sill2015emerging, burns2016borg, k8s}. Kubernetes is a Google open project advocating the vision of a modular, customizable and therefore scalable orchestration platform \cite{bernstein2014containers}. \textcolor{black}{In order to guarantee the availability and continuity of hosted applications, Kubernetes uses the Raft protocol. The latter replicates the states between the machines hosting the containers, where each state is an image of the hosted containerized applications 
} 
 \cite{oliveira2016evaluating,ongaro2014search}. In spite of its simplicity and rapidity in the replication process, Raft protocol has major limitations when it comes to machines' failure. Indeed, Raft can only detect and correctly deal with shutdown events of machines. In other words, if a machine experiences a Byzantine (random) behavior, Raft is unable to guarantee service continuity \textcolor{black}{\cite{ongaro2014search,sousa2013state,tangora}. Indeed,  
 Byzantine behaviors, such as delayed, dropped, or corrupted messages, or abnormally executed processes have been widely observed in real systems, as summarized in \cite{Correia2012}}.
 

Being conscious of the risks of software errors and malicious attacks that can push a machine into a Byzantine behavior, we propose in this paper the adaptation and integration of the BFT-SMaRt fault-tolerant replication protocol into Kubernetes. By doing so, we expect our proposed platform to resist to any type of faults while guaranteeing service continuity. To the best of our knowledge, this is the first work that proposes a Kubernetes platform tolerant to Byzantine and non-Byzantine faults. 

The main contributions of this paper are summarized as follows:
\begin{enumerate}
\item We present an overview of Docker virtualization, Kubernetes platform, fault-tolerance within this platform and its limits.
\item We propose the Kubernetes multi-Master Robust (KmMR) platform, a platform tolerant to Byzantine and non-Byzantine faults. KmMR is based on the integration of BFT-SMaRt into Kubernetes environment.
\item We propose an efficient method to adapt and integrate the replication protocol BFT-SMaRt (written in Java) into Kubernetes (written in Golang). 
\item We implement the proposed KmMR solution 
in an OpenStack-based cloud environment, evaluate its performances and compare it to the conventional platform, called Kubernetetes multi-Master Conventional (KmMC). Comparison is realized through experiments in non-Byzantine and Byzatine environments, where both crash and Distributed Denial-of-Service (DDoS) attacks are performed to destabilize the machines and corrupt their replication process. 
The obtained results confirm the effectiveness and robustness of KmMR.
\end{enumerate}
The rest of the paper is organized as follows.
Section II describes Docker containerization technology. In section III, the Docker containers orchestration platform Kubernetes is explained. Whereas, section IV discusses fault-tolerance in Kubernetes. Section V presents our KmMR platform. Experimental evaluation and results are discussed in Section VI. Finally, section VII closes the paper.  
\section{Background} 
In this section, we present an overview of Docker containers and its orchestration \textcolor{black}{mechanism, supported by Kubernetes}. 

\subsection{Docker Containers} 
\begin{figure}[t]  
  \centering
  \includegraphics[width=340pt]{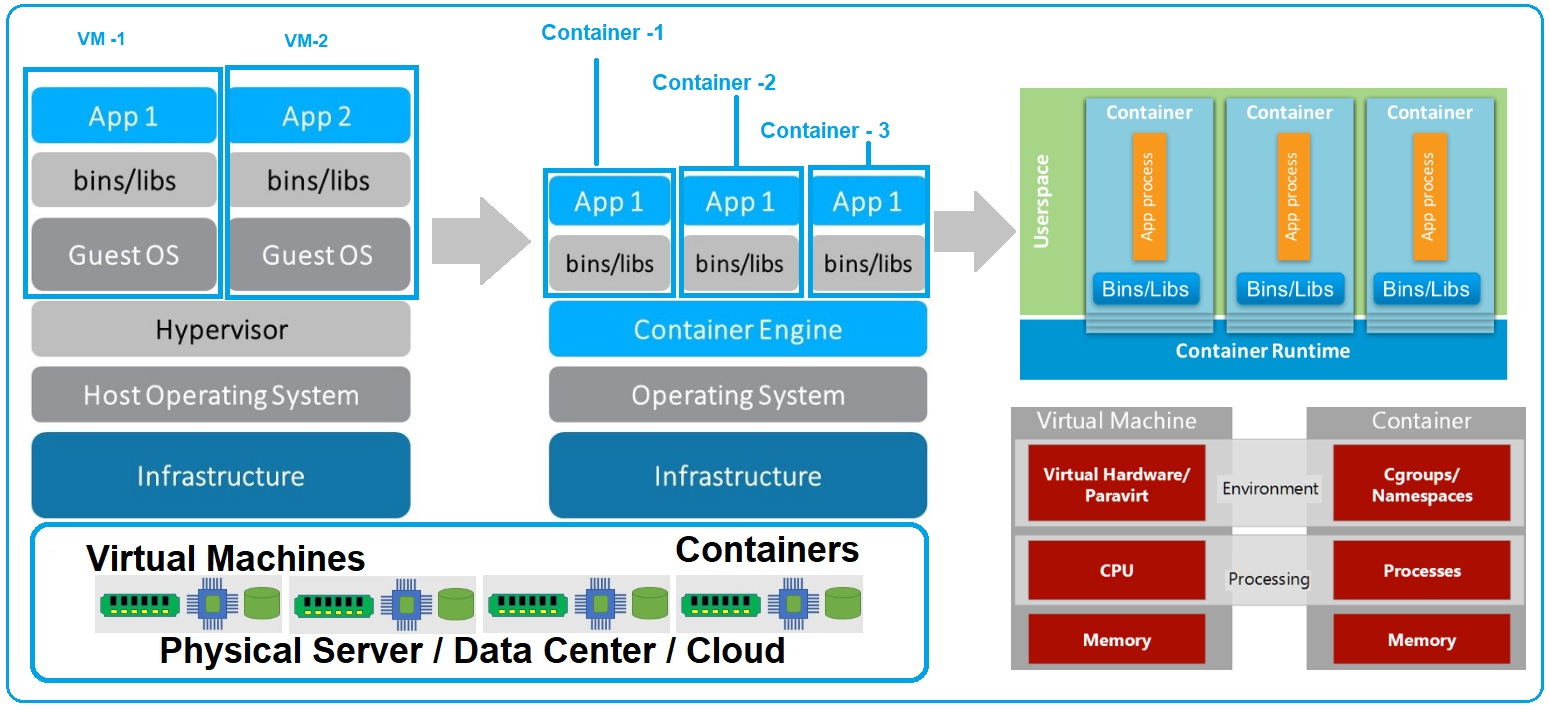}
  \caption{From virtual machines to containers \cite{Prasad}}
  \label{VM_Cont}
\end{figure}

Container virtualization, also known as containerization, relies directly on kernel functionalities to create isolated virtual environments, as illustrated in Fig. \ref{VM_Cont}. These virtual environments are named containers, while the features provided by the operating system 
kernel are called 
\textit{namespaces} and \textit{cgroups} \cite{moga2016level}. The \textit{namespaces} control and limit the amount of resources used for a process, while the {\textit{cgroups}} manage the resources of a process group. Hence, a container provides the resources needed to run applications as if they were the only processes running in the host machine's operating system. 

\textcolor{black}{Even though several containerization platforms have been proposed, such as LXC, OpenVZ and Docker \cite{bernstein2014containers,peinl2016docker,Rizki}, only Docker sparked interest and popularity among the research and professional communities thanks to its operational simplicity and flexibility.} Indeed, traditional virtualization uses a Hypervisor to create virtual machines, \textcolor{black}{deploy guest operating systems on them, and host the applications \cite{VMwarevSphere}}. \textcolor{black}{Whereas}, Docker containerization requires only the installation of the Docker \textcolor{black}{container engine} on the operating system's kernel of the host machine, \textcolor{black}{allowing the creation of containers that host the applications}. Nevertheless, both are autonomous systems that use a higher system, \textcolor{black}{i.e., the one} of the host machine, to perform their tasks. The difference is that virtual machines must contain a whole \textcolor{black}{guest operating system, while the containers use directly the one of the host machine. Fig. \ref{VM_Cont} illustrates an architectural comparison of traditional virtualization and containerization.}



\begin{figure}[t]
\centering
\includegraphics[width=220pt]{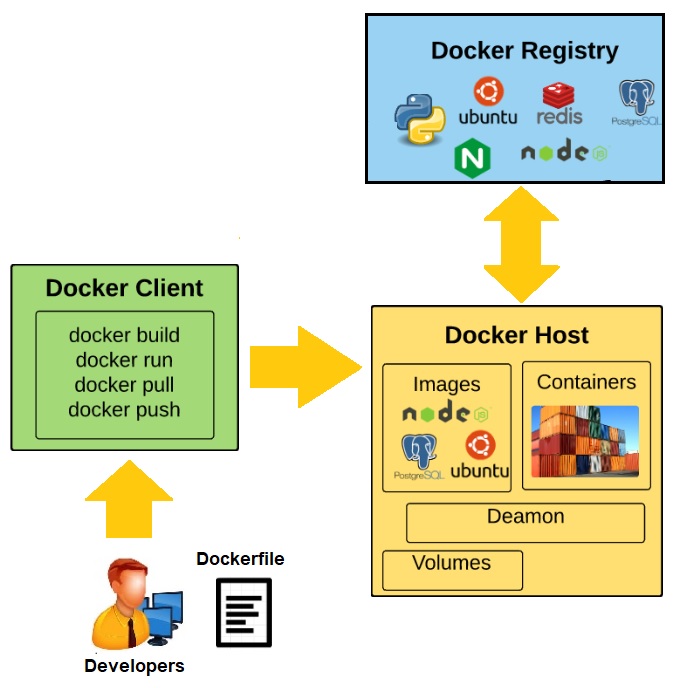}
\caption{Components of Docker \cite{Prasad}}
\label{architectureDocker}
\end{figure}

\textcolor{black}{Docker is a complex but very intuitive ecosystem for container development. It is mainly composed of six elements, as shown in Fig. \ref{architectureDocker}:
\begin{enumerate}
    \item \textit{Docker Client:} It is the command line interface tool used to configure and interact with Docker. For any Docker command instruction (e.g., $docker \; run$), the client sends the command to Docker daemon ($dockerd$) that carries it out. 
    \item \textit{Docker Daemon:} It is the Docker server that listens to the application programming interface requests and manages Docker objects, such as images, containers, networks, and volumes, etc.
    \item \textit{Docker Images:} An image is a read-only template/snapshot, pulled or pushed from a public or a private repository, in order to create a Docker container. This is the building block of Docker. It is lightweight, small, and fast compared to those of traditional virtual machines.
    \item \textit{Docker file:} It is used to build Docker images.
    \item \textit{Docker Containers:}
    Basically, a Docker container is a user space of the operating system. It is composed of a set of processes isolated from the rest of the system, and running from an image that provides necessary files to support the processes of the hosted application.
    \item \textit{Docker registries:} Registries are the central repository and distribution component of Docker images. 
    \item \textit{Docker Engine:} It combines the Docker daemon, application programming interface and the command line interface tools.
\end{enumerate}
}

\textcolor{black}{By its simplicity and small number of components, the Docker architecture provides interesting advantages \cite{Rad2017AnIT,joy2015performance,Vase}: \begin{enumerate}
    \item \textit{Deployment rapidity:} Docker achieves fast operations, such as communication, and container building, testing and deployment.
    \item \textit{Applications Portability:} Containerized applications are easily portable, as they can be moved around as a single unit, without affecting their response performances or the containers.
    \item \textit{Fast service delivery:} Docker containers format is standardized, such that programmers and administrators tasks do not interfere when deploying them. Indeed, Docker provides a reliable, consistent, and enhanced environment that  achieves predictable outputs when codes are moved between development, test and deployment platforms.  
    \item \textit{Density:} Docker uses the available resources more efficiently compared to virtual machines, since it does not rely on a Hypervisor. It is able to run densely several containers on the same single host, hence optimally using the resources and increasing its performance, compared to virtual machines.
    \item \textit{Scalability:} Docker can be deployed in several physical servers, data servers, and cloud platforms without any restriction. Containers can be easily moved from a cloud environment to a local host and vis-versa, at a fast pace. Deployment adjustments can be easily realized according to needs.
\end{enumerate}}

\begin{table}
\caption{Characteristics of Virtualization Technologies}
\label{table:conceptsVirtuals}
\begin{tabular}{|m{2.5cm} | m{4.16cm}| m{4.16cm} | } 
\hline
\rowcolor[gray]{0.75} 
\textbf{Parameters} & \textbf{Virtual Machines}  & \textbf{Docker Containers} \\
\hline
\textit{Operating System} & Every virtual machine virtualizes the host material and loads its own guest operating system & No container emulates host material. Host operating system is used. \\
\hline
\textit{Communication} & Through Ethernet peripherals & Through Inter-Process Communication standard mechanisms, e.g., sockets, pipes, shared memory, etc. \\
\hline
\textit{Resources Usage (CPU and RAM)} & High  &  Quasi-native\\
\hline
\textit{Startup time} & Few minutes & Few seconds  \\
\hline
\textit{Storage} & High requirement for guest operating system and associated software installation and execution  &
Low since host operating system is used \\
\hline
\textit{Isolation} & Libraries and files' sharing among virtual machines is impossible & Libraries and files can be seamlessly mounted and shared \\
\hline
\textit{Security} & Depends on the Hypervisor's configuration & Requires access control \\
\hline
\end{tabular}
\end{table}

In Table \ref{table:conceptsVirtuals}, we summarize the characteristics of {virtual machines} and Docker containers. {Accordingly}, the {Docker} container can be created and removed almost in real time and thus introduces a negligible task overload with respect to the host machine's resources use \cite{ joy2015performance,felter2015updated,sharma2016containers}. Compared to \textcolor{black}{virtual machines, Docker} containers are advantageous in terms of network management, boot speed, deployment/migration flexibility, and resources use, {e.g. RAM, storage, etc.} \cite{joy2015performance}. However, they suffer from the weak isolation of the host machine. Indeed, if a Docker container is compromised, then an attacker can get full access to the operating system of the host machine \cite{7092943,manu2016docker}. Consequently, there is an urgent need for a robust and secured environment for Docker containers. Moreover, Docker is unable on its own to deploy containers on distributed machines and ensure their interaction \cite{7092943}. In this matter, an orchestration mechanism is needed to manage Docker containers in distributed systems. 

\subsection{Containers Orchestration} 
Handling a few Docker containers on one machine is an easy task. However, when it comes to moving these containers into production on a set of distributed hosts, many questions arise. Indeed, driven by providing availability, scaling, and networking, an integration and management tool is required not only {to ensure} initial containers deployment, but {to} also {manage} multiple {and dynamic} containers as one entity. Clearly, handling everything manually is not conceivable because it would be very difficult to ensure the viability, maintenance and sustainability of the system. Thus, the process of deploying multiple containers can be optimized through automation, especially {in large scale systems}. This type of automation is referred to as orchestration and includes features like work nodes’ location determination, load balancing, inter-container communication, service discovery, {status} updates, {containers} migrations, scaling up, and tolerance to malfunctions.

Several orchestrators have been proposed and implemented {to manage Docker container-based platforms}. Examples include Fleet \cite{fleet}, Mesos \cite{mesos}, Swarm \cite{luzzardiswarm} and Kubernetes \cite{kubernetes}. In the remainder of this paper, we are interested in Kubernetes only. The latter is a stable and free solution that can automate the deployment, migration, monitoring, networking, scalability, and availability of applications \textcolor{black}{hosted in container-based server platforms} \cite{peinl2016docker, burns2016borg}. 
\section{Kubernetes: An Open-Access Orchestrator of Docker Containers}
Kubernetes, abbreviated K8s, is a project initiated by Google in 2014 when it saw the advantages of Docker containers over traditional virtualization. 
The Kubernetes Orchestrator automates the deployment and management of large-scale containerized applications, \textcolor{black}{such as applications' microservices generation \cite{microserv}, cloud services to store, access, edit and share video content \cite{GoPro}, and mission critical services as telecommunications and energy delivery services \cite{Mazzara,Baner}}. Its platform runs and coordinates containers on sets of physical and/or virtual machines. Kubernetes is designed to fully manage the life cycle of containerized applications, using predictability, extensibility, and high availability methods, as detailed in \textcolor{black} {\cite{kratzke2017understanding,K8sMaster}}. 

\begin{figure*}
  \centering
  \includegraphics[width=340pt]{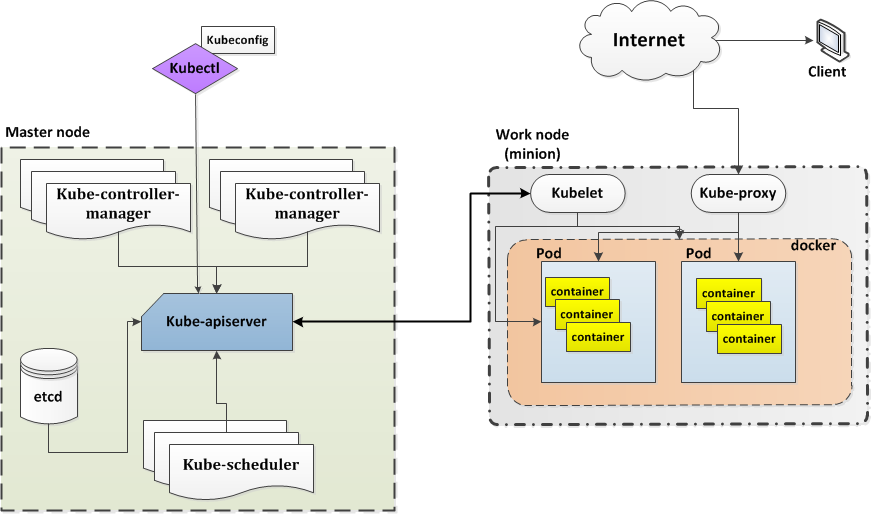}
  \caption{Architecture of Kubernetes (ex: one master node and one work node)}
  \label{Kube_archi}
\end{figure*}

\subsection{Kubernetes Architecture}
Kubernetes architecture is based on the master/slave model \cite{bila2017leveraging}. It consists of a cluster of one master node and several work nodes, called $minions$, as shown in Fig. \ref{Kube_archi}. Their roles are given as follows:\\
\textit{\textbf{Kubernetes master}}: This node is responsible of the overall management and availability of the Kubernetes cluster. Its components, i.e. \textcolor{black}{the Application Programming Interface} (API) server,
controller and scheduler, support the interaction, monitoring and scheduling tasks
within the cluster. The API server provides the interface to the shared state of the cluster through which the other components, e.g. work nodes, interact. The controller monitors the shared state of the cluster through the API server and makes decisions to bring the cluster back from an unstable state to a stable one. The scheduler manages the cluster load. It takes into account individual and collective resource requirements, quality-of-service requirements, hardware/software constraints, policies, etc. The Kubernetes cluster data is stored in a database, e.g. etcd \cite{etcd}, whereas cluster administration is at the master level via the K8s command-line interface $kubectl$. The latter stores its configuration and authentication information to access the API server in the $kubeconfig$ file.\\
\textit{\textbf{Kubernetes minions}}: Containerized applications run on these nodes. On one hand, the client nodes communicate with the work node via their $kubelet$ through the master node. The $kubelet$ receives commands from the master node and executes them through its Docker engine. It also reports the state of the work node to the API server. On the other hand, the \textit{kube-proxy} runs on each work node to manage clients' access to deployed services. Each service is compiled into one or many \textit{Pods}. A \textit{Pod} is a logical set of one or several containers. This is the smallest unit that can be programmed as a deployment in Kubernetes. Containers in the same \textit{Pod} share resources such as storage capacity, IP address, etc.
\subsection{Pods Instantiation}
In Kubernetes, the placement of \textit{Pods} is realized following a specific strategy. 
In fact, considering a Kubernetes cluster consisting of a master node and a finite set of \textit{minions}  $\mathcal{M} = \{M_1, M_2, ..., M_n \} $, a pod $P(t, m, p, v)$ asking for $t$ CPU cycles, $m$ RAM, a specific communication port $p$ and a $v$ storage capacity, needs to be deployed within the cluster. To select the $minion$ on which the pod will be instantiated, the K8s master node proceeds in two steps: 1) it filters the \textit{minions}. Then, 2) it ranks the remaining \textit{minions} to determine the best one suited for the pod. These two steps are detailed as follows:\\
\textit{\textbf{Filtering}}: In this operation, nodes without required resources $(t, m, p, v)$ are removed. Kubernetes uses multiple predicates to perform filtering, including:
    \begin{itemize}
    \item \textit{PodFitsResources}: does the node have enough resources (CPU and RAM) to accommodate the pod?
    \item \textit{PodFitsHostPorts}: is the node able to run the pod via the $p$ port without conflicts?
    \item \textit{NoVolumeZoneConflict}: does the node have the amount of $v$ storage that the pod requests?
    \item \textit{MatchNodeSelector}: does the node match the parameters of the selector query defined in the pod description?
 \end{itemize}
These predicates can be combined to set up sophisticated filters.\\
    \textit{\textbf{Ranking}}: After filtering, Kubernetes uses priority functions to determine the best $minion$ among the nodes able to host the pod. A priority function assigns a score between 0 and 10 where 0 is the least preferred and 10 is the most preferred node. Each priority function is weighted by a positive number and the final score is the sum of the weighted scores. 
    The main priority functions that can be activated in Kubernetes are:
    \begin{itemize}
    \item \textit{BalancedResourceAllocation}: it aims at balancing the $minions$ charge. Indeed, it places the pod in a node in a way that the resource utilization rate (CPU and RAM) is balanced among the minions.
    \item \textit{LeastRequestedPriority}: it favors the node that has most resources available.
    \item \textit{CalculateSpreadPriority}: it minimizes the number of pods belonging to the same service on the same node.
    \item \textit{CalculateAntiAffinityPriority}: it minimizes the number of pods belonging to the same service on nodes sharing a particular attribute or label. 
    \item \textit{CalculateNodeLabelPriority}: it favors nodes with a specific label.
    \end{itemize}
Once the final scores of all nodes are calculated, the $minion$ having the highest score is selecte to instantiate the pod. If there is more than one $minion$ that has the highest score, the master node selects one of them randomly.
\section{Fault Tolerance in Kubernetes}
In this section, we explain the fault tolerance mechanism in Kubernetes. We start by a brief description of faults. Next, we present the associated consensus problem. Finally, the built-in fault tolerance protocol ``Raft" is detailed.
\subsection{Background}
The robustness of a system refers to its ability to continue functioning when part of the system fails \cite{cristian1991understanding}. A system fails when the outputs are no longer conform to the original specification. 
The occurrence of a failure can be: 1) transient, i.e. appears, disappears and never occur again, 2) intermittent, i.e. reproducible in a given context and 3) persistent, i.e. appears until repair. A non-faulty (non-failing) node or process is called correct when it follows its specifications. Whereas, a faulty node/process may stop or exhibits a random behavior. In general, failures/faults may be caused by software defects, malicious attacks, or human-machine interaction errors.
In distributed systems orchestrated by Kubernetes, faults may occur at the master node or minions. 
They can be classified into two categories:
\begin{enumerate}
    \item \textit{Fail-stop faults}: They are characterized by the complete activity's stop (or crash) of a node. This state is perceived by others as the absence of expected messages until the eventual application's termination. A system that is able to detect only these faults considers that a node/process can be in one of two states, either it works and gives the correct result, or it does nothing.
    \item  \textit{Byzantine faults}: Byzantine faults are characterized by any behavior deviating from the node/process's specifications and producing non-conform results \cite{lamport1982byzantine}.
    We distinguish between natural Byzantine faults, such as undetected physical errors on messages' transmissions, memory and instructions, and malicious Byzantine faults, designed to defeat the system, such as viruses, worms and sabotage instructions.
\end{enumerate}
In large and/or uncontrolled systems, the risk of faults is high and shall be mitigated to ensure service continuity. 
One way to realize it is to use the State Machine Replication (SMR) mechanism \cite{aublin2014vers}. The latter consists of using multiple copies of a system, implemented as a state machine, to tolerate faults and keep the system's availability. Each copy of the system, called a replica, is placed on a different node \cite{schneider1990implementing}.
SMR allows a set of nodes to execute the same instruction sequences on each request sent by a client. There are two approaches to execute requests: 1) active replication, where all nodes execute requests, update their state machines, and respond to clients. And 2) passive replication, where only one node, called \textit{leader}, executes the requests and forwards state machine changes to other nodes, then responds to clients.\\
To avoid inconsistency in replication, nodes/replicas need to be sure that their state machines are identical before responding to clients. The following section describes this state machine replication problem, called the \textit{Consensus} problem.
\subsection{Consensus Problem}
The \textit{Consensus} is a fundamental condition in fault-tolerant distributed systems. It consists of tuning replicas' values to the same one, proposed by one of the nodes. The \textit{Consensus} problem can be formulated as follows: We assume a system composed of a set $\mathcal{N} = \{N_1, N_2, \ldots, N_{n}\}$ of $n$ replicated nodes, and that at most only $f$ nodes can fail, where $f \leq n-1$. Let $\mathcal{N}' \subseteq \mathcal{N}$ be a subset of $m \leq n$ nodes. The consensus problem consists of finding a protocol that allows the following:
\begin{enumerate}
         \item Any node $\in \mathcal{N}'$ can propose a replica's value to the other nodes.
         \item When all nodes agree on the same value, a consensus is achieved.
\end{enumerate}
Without loss of generality, protocols that satisfy these conditions, possess four properties \cite{pease1980reaching}: 
    \begin{enumerate}
    \item \textit{Termination}: Each correct node eventually decides a value.
    \item \textit{Validity}: The decided value has been proposed by one or many other nodes.
    \item \textit{Integrity}: The decision is unique and final.
    \item \textit{Agreement}: Two correct nodes cannot decide different values.
\end{enumerate}
According to \cite{schneider1990implementing}, any protocol that verifies the following safety and liveness conditions has the previous four properties:
\begin{enumerate}
\item \textit{Safety}: All the correct replicas execute the requests they receive in the same order.
\item \textit{Liveness}: Each request is correctly executed by correct nodes.
\end{enumerate}
Such a protocol is commonly referred as consensus/replication protocol. Its decisions are based on exchanged messages between all or a part of the nodes in the system. Indeed, a consensus is achieved if the quorom, defined as the minimum number of correct nodes required to build the consensus, participate in the consensus process. The quorum depends on the size of the system and the maximum number of tolerated faults.\\
Two fault-tolerant classes of replication protocols exist. In the first, called \textit{Non-Byzantine}, nodes fail only when they stop functioning. For $n$ nodes, at most $f=\frac{n-1}{2}$ crash faults can be tolerated. Examples of non-Byzantine protocols include Raft \cite{ongaro2014search}, Paxos \cite{lamport2001paxos}, and Zab \cite{van2015vive}. In the second class, called \textit{Byzantine}, any type of failures can be tolerated. However, they typically tolerate only $f'=\frac{n-1}{3}$ faults \cite{bracha1985asynchronous}. As Byzantine protocols examples, we can cite Practical Byzantine Fault-Tolerance (PBFT) \cite{castro2002practical}, Efficient Byzantine Fault-Tolerance (EBFT) \cite{veronese2013efficient}, UpRight \cite{clement2009upright}, Prime\cite{amir2011prime}, and Byzantine Fault-Tolerance State Machine Replication (BFT-SMaRt) \cite{sousa2013state, sousa2018byzantine}.   
\subsection{Built-in Fault Tolerance in Kubernetes: Raft Protocol}
\begin{figure}[t]
  \centering
   \includegraphics[width=250pt]{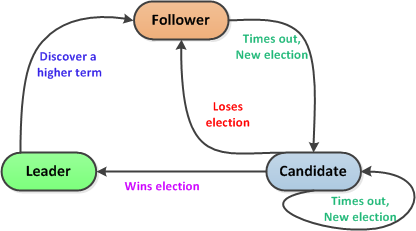}
  \caption{Raft Protocol's Election Process}
  \label{fig:leaderRaft}
\end{figure}

Raft is the replication protocol built into Kubernetes \cite{ongaro2014search,raftsite}. Basically, it ensures that the replicas maintain identical state machines, while tolerating only crash faults. It is based on passive replication, where a node may be \textit{leader}, \textit{follower} or \textit{candidate}, as illustrated in Fig. \ref{fig:leaderRaft}:
\begin{itemize}
    \item \textit{Leader}: In a cluster, a single active node directs the communication, by receiving requests, processing them, forwarding state machine changes to other nodes, and responding to clients.
    \item \textit{Follower}: When a \textit{leader} is active, all other nodes are set as \textit{followers}. They wait for the changes sent by the \textit{leader} to update their state machines. 
    \item  \textit{Candidate}: When the \textit{leader} breaks down, the \textit{followers} become \textit{candidates} and trigger votes to elect a new \textit{leader}.
\end{itemize}  
The \textit{mandate} of a \textit{leader} lasts from its election until its breakdown. In order to organize elections, Raft assigns an index to each mandate. These indexes are called \textit{terms}. Any leader or candidate node includes the \textit{term} index in its messages. Whereas, a \textit{follower} needs to wait for a random time, typically between $ 150$ and $ 300$ ms, before transiting into \textit{candidate}. An active \textit{leader} periodically sends heartbeat messages ($ AppendEntriesMessage$) to all nodes in the cluster. Any node receiving this message resets its  wait time to a random value. Otherwise, at the expiration of its wait timer, the \textit{follower} changes status to \textit{candidate} and triggers a new election. The \textit{candidate} proceeds as follows: 1) Increments its current term number, 2)  votes for itself, and 3) sends vote request ($RequestVoteMessages$) to all other nodes.
The latter vote for the request containing a \textit{term} index greater than theirs, update their term index and return to the \textit{follower} status. Once a candidate receives the votes of the majority, defined as $ \lceil {f + 1} \rceil$ votes, it becomes the new \textit{leader}. However, if no candidate obtains the majority of votes, e.g. in a tie situation, no leader is elected in this term, and a new term will be triggered by the node that sees its timer expiring first.
The requirement for a majority of votes ensures that a single \textit{leader} is elected in a \textit{term} (\textit{Safety} condition), while the wait time of \textit{followers} guarantees that a leader will eventually be elected (\textit{Liveliness} condition). 

To run in Kubernetes environment, some changes have been made to Raft protocol:
\begin{enumerate}
    \item Unlike the conventional Raft, where requests to \textit{followers} are redirected to the \textit{leader}, Raft is converted to active replication to be conform to the load balancing property of Kubernetes \cite{raftsite}. 
    \item Raft is re-implemented in Golang, the same programming language used to develop Kubernetes and Docker containers.
\end{enumerate}
Besides Raft, another non-Byzantine replication protocol, called $DORADO$ was proposed for Kuberenetes \cite{netto2017state}. This protocol is similar to Raft, but requires sharing the master node's memory to all instanciated containers in work nodes, in order to store their state machines. This approach allows to achieve shorter consensus times than Raft, but aggravates the containers' isolation issue.

Despite their simplicity, Raft and $DORADO$ are particularly powerless against Byzantine behaviors \cite{lim2014scalable}. Indeed, a failing node may not stop, and adopts continually a Byzantine (random) behavior, e.g. not following the protocol, corrupting its local state, or producing incorrect or inconsistent outputs \cite{schneider1990implementing}. 
To mitigate this problem, we propose in the next section a novel Kubernetes platform, where both non-Byzantine and Byzantine faults can be tolerated, while ensuring service continuity.

\section{KmMR: A K8s multi-Master Robust Platform}  
Kubernetes allows to deploy and orchestrate groups of containers with a single master node. The latter replicates the containers on different work nodes to provide service continuity. However, if the master node fails, containers are no longer available and all management data is lost. To avoid such case, the deployment of multi-master clusters, where several master nodes cooperate, becomes necessary. {However}, duplicating master nodes only does not provide complete fault tolerance \cite{perronne2016vers}. In fact, {this mechanism} must be associated with a replication protocol to ensure consistency between the master nodes states, \textcolor{black}{i.e., update operations to a replicated data item within the nodes should reach and be executed at some time, in all master nodes, and in the same chronological order \cite{schneider1990implementing,Souri}.
Such} multi-master systems are important for critical applications, {e.g.,} telecommunication and energy services, where the continuous availability of services is required 24 hours a day, and 7 days a week.

\begin{figure}  
  \centering
  \includegraphics[width=270pt]{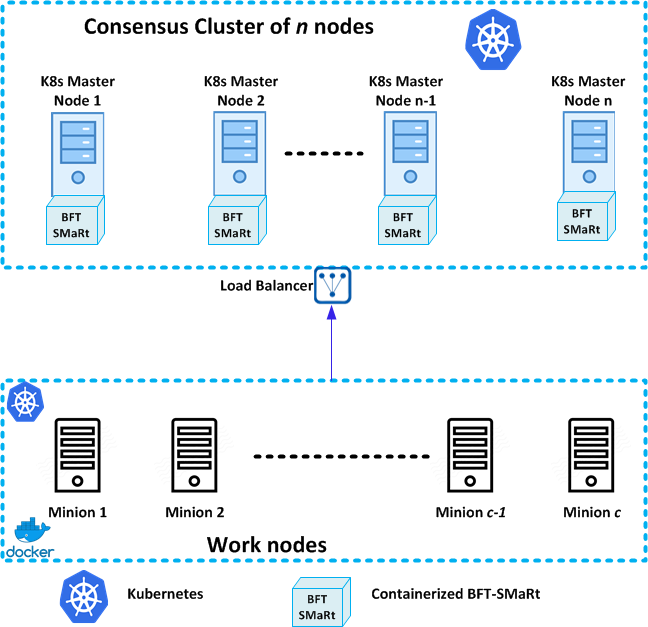}
  \caption{System Model}
  \label{plateforme}
\end{figure}

In this section, we propose to create a resistant Kubernetes multi-master platform to all kinds of faults, in order to guarantee service continuity. We consider a Kubernetes cluster consisting of $n$ replicated K8s master nodes and $c$ work nodes. Work nodes process clients' service requests and send their reports (requests) to the master nodes, as shown in Fig. \ref{plateforme}. We assume that communications between nodes may experience important delays, thus causing communication failures.
\begin{figure}  
  \centering
  \includegraphics[width=250pt]{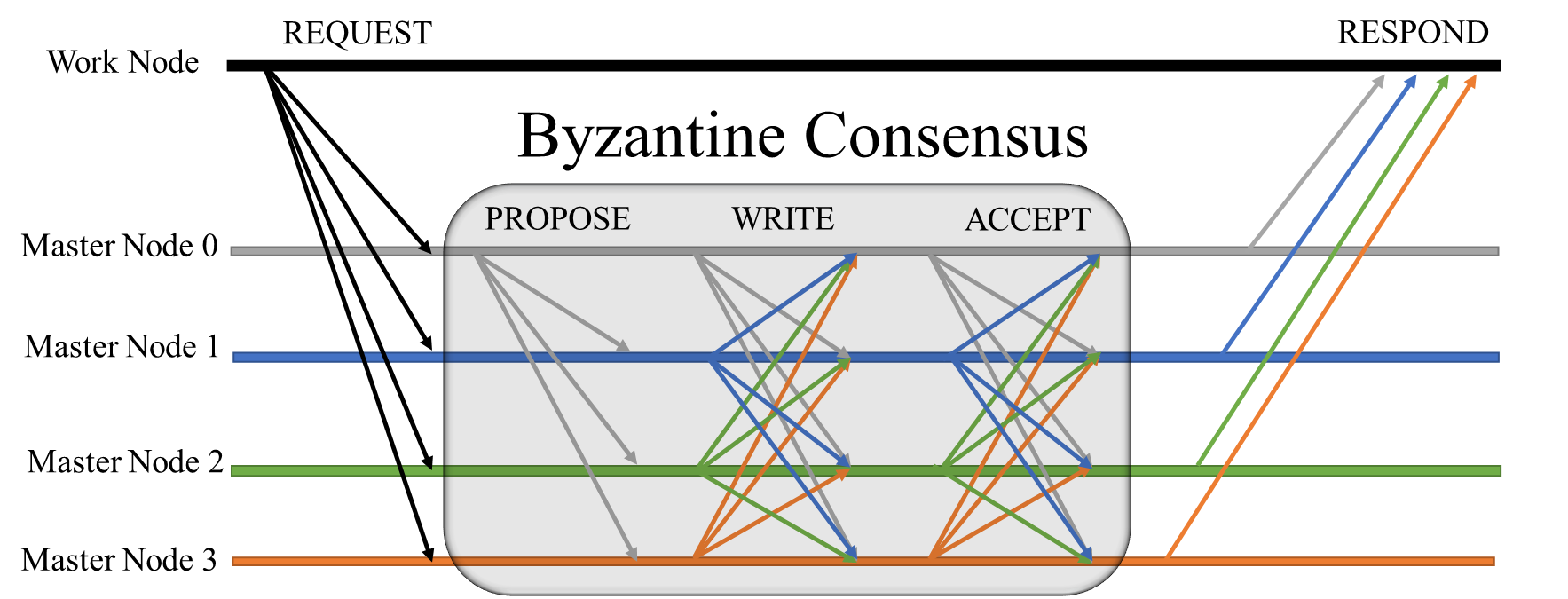}
  \caption{Consensus Process by BFT-SMaRt}
  \label{bftsmart}
\end{figure}



\subsection{BFT-SMaRt: Replication Protocol for KmMR}
Among the known Byzantine protocols, only PBFT \cite{castro2002practical}, UpRight \cite{clement2009upright} and BFT-SMaRt \cite{sousa2013state} implement a Byzantine fault-tolerant replication system. The choice of BFT-SMaRt is motivated by the following:
\begin{itemize}
\item BFT-SMaRt is very well suited for modern hardware, e.g. multi-core systems, unlike other protocols such as PBFT {\cite{sousa2013state}}.
\item BFT-SMaRt outperforms other protocols, e.g. UpRight, in terms of consensus time, defined as the required time to process a client's request {\cite{sousa2013state}}. 
\item BFT-SMaRt guarantees a high accuracy in replicated data, when a Byzantine faulty behavior is exhibited within the system {\cite{sousa2013state}}.
{\item BFT-SMaRt is a modular, extensible and robust library. It is able to provide an adaptable library that sets-up reliable services \cite{BFT-SMaRtlibrary}.} 
\item Unlike other Byzantine protocols, BFT-SMaRt supports reconfiguration of the replica sets, e.g., addition and removal of nodes {\cite{lamport2010reconfiguring}}.
{\item BFT-SMaRt provides efficient and transparent support for critical and sustainable services \cite{bessani2013efficiency}.}
\end{itemize}
In BFT-SMaRt, a consensus is established according to the following steps, as illustrated in Fig. \ref{bftsmart}.  
First, a work node broadcasts its request to master nodes, who trigger the execution of the consensus protocol. Each instance of the consensus begins with the \textit{leader} master node proposing to other nodes a batch of requests in the \textit{PROPOSE} message. Master nodes validate the authenticity of the \textit{PROPOSE} message and its content. If valid, they register the proposed batch and broadcast \textit{WRITE} messages with cryptographic hashes of the proposed batch, to all other nodes. 
 If a master node receives $ \lceil \frac{n + f' + 1}{2} \rceil$ \textit{WRITE} messages with the same hash, \textcolor{black}{where $\lceil.\rceil$ is the ceiling function,} it sends an \textit{ACCEPT} message to all other nodes. This message contains its decision batch for the consensus instance.
If the \textit{leader} master node is not correct, a new election must be triggered, and all nodes need to converge to the same execution by consensus. \textcolor{black}{The election} procedure is described in detail in \cite{sousa2012byzantine}.
\begin{figure*}  
  \centering
  \includegraphics[width=350pt]{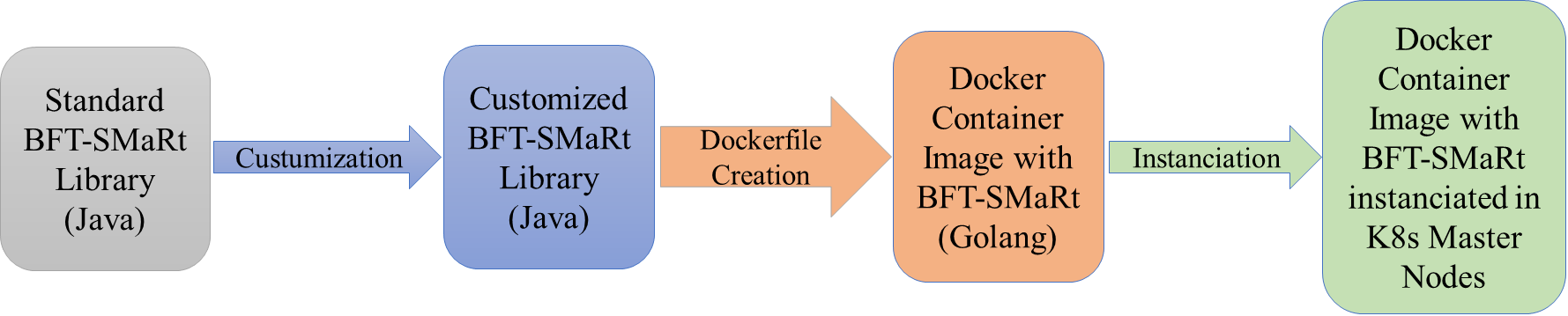}
  \caption{Integration Methodology of BFT-SMaRt into Kubernetes}
  \label{integrationBftK8s}
\end{figure*}
\subsection{Proposed Integration Methodology of BFT-SMaRt into K8s}
\begin{figure}  
  \centering
  \includegraphics[width=250pt]{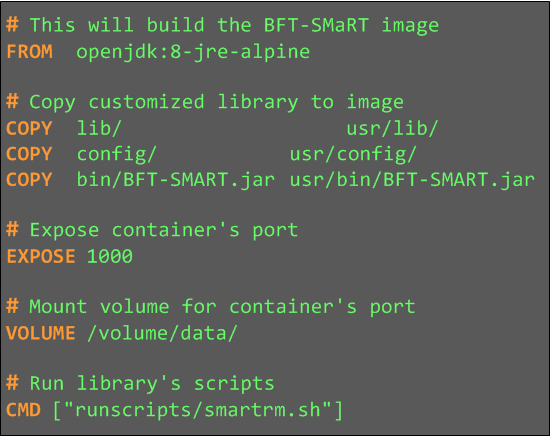}
  \caption{Dockerfile to Create the BFT-SMaRt Container}
  \label{Dockerfile}
\end{figure}

The BFT-SMaRt protocol is implemented in Java, an object-oriented programming language, while Kubernetes and the Docker engine are written in Golang, a service-oriented programming language \cite{golang}. In order to integrate BFT-SMaRt into Kubernetes, two options can be considered: 
\begin{enumerate}
    \item Rewrite all BFT-SMaRt library's source code in Golang.
    \item Wrap the BFT-SMaRt library in a Docker container.
\end{enumerate}
Unlike Raft, with a source code less than 3000 lines and easily rewrited in Golang, BFT-SMaRt source code is larger and more complex, with approximately 100 files and a total of 13500 lines of Java code. Consequently, the second option is more likely to be realizable. This choice is supported by the advantages offered by Docker. Indeed, Docker containers run fast and their introduced overhead is negligible \cite{felter2015updated,joy2015performance}.
The proposed procedure to integrate BFT-SMaRt into Kubernetes is illustrated in Fig. \ref{integrationBftK8s}. First, we recover the library BFT-SMaRt and all its dependencies from Github \cite{BFT-SMaRtlibrary}. Then, we customize it by setting the parameters of the master nodes. Next, we create our Docker file \textit{Dockerfile}, as detailed in Fig. \ref{Dockerfile}. Afterwards, we execute \textit{Dockerfile} to produce the BFT-SMaRt containerized image. Finlly, we instantiate in each K8s master node the Docker image with its information.
\section{Experimental Evaluation} 
\subsection{Simulation Settings} 
We implemented the KmMR solution in an OpenStack cloud environment provided by Ericsson Canada \cite{sefraoui2012openstack}. The available resources are as follows: 50 GB of RAM and 20 virtual processors (VCPU), usable on a maximum of 10 machines.

The experiment is carried out on clusters composed of several Kubernetes master nodes ($n=5$ and $n=7$), connected to each other via the OpenStack GigabitEthernet network and accessible from the Internet. Each node is a virtual machine equipped with the Ubuntu server 18.04 TLS 64-bit OS, a dual-core i7 CPUs (VCPU) clocked at 2.4 GHz, 4 GB of RAM and 20 GB storage capacity. The Docker engine 18.05.0-ce is installed on Kubernetes nodes for container instantiation needs. We deployed Kubernetes 1.11.0 to orchestrate the Docker containers. The master Kubernetes role \textit{kubeadm} has been enabled on all master nodes (multi-master configuration). The remaining machines are used to act as work nodes and DDoS attackers. BFT-SMaRt has been containerized and integrated into the master nodes to provide coordination and consensus. Work nodes send their requests in closed loop, i.e. they wait for the response of a request before sending a new one, as defined in \cite{schroeder2006open}. 

In the cluster, we initialize the replication protocol on master nodes. Then, two work nodes broadcast their requests. Upon request reception, master nodes exchange messages to build the consensus. To measure the performance of KmMR, we used the micro-benchmark $0/0$ where both request and response messages are empty \cite{castro2002practical}. 



\textcolor{black}{DDoS attacks are used to model Byzantine behaviours, using the \textit{Hping3} command \cite{hping3, ops2016denial}. Indeed, we inject DDoS-based ``CPU Load" and ``Network Flooding" Byzantine faults as follows \cite{Gupta2016_2,Gupta2016}. ``CPU Load" fault is triggered by increasing the number of users continuously sending requests to a master node, while ``Network Flooding" can be initiated by some master nodes towards others. We assume that}
attacking machines target simultaneously a single master node. Each attacker sends successively and continuously requests of size 65495 bytes in open loop, i.e. without waiting for responses, through the command \textit{Hping3 -f IP address of targeted master node -d 65495}. \\
We evaluate the performance of our solution and compare it to the Kubernetetes multi-Master Conventional (KmMC) platform, where non-Byzantine replication protocol Raft is used. Two scenarios are considered for our experiments:
\begin{itemize}
 \item \textit{Scenario 1}: In this scenario, we consider a
 Kubernetes platform where, initially, the number of (crash) faults in the cluster is lower than the maximum number of faults tolerated by the replication protocol in place. This corresponds to $f < \frac {n - 1} {2}$ and $f' < \frac {n - 1} {3}$ for KmMC and KmMR respectively. Then, we perform a DDoS attack on one master node, and evaluate the consensus times for each platform. 
\item \textit{Scenario 2}: Unlike \textit{Scenario 1}, the initial number of (crash) faults is set to be the maximum that can be tolerated by the used replication protocol. Then, DDoS attacks are performed on one master node. In this scenario, we evaluate established consensus times as well as resources consumption by the DDoS victim (CPU, RAM, and available communication Bandwidth). Resources are measured using commands \textit{IPerf3} for Bandwidth, and \textit{top} for CPU and RAM \cite{iperf3,top}. 
\end{itemize}
\subsection{Results and Discussions} 
\begin{table*}
\begin {center}
\caption{Consensus Times ($\mu$sec) versus DDOS Attack Rate ($Gbps$) (\textit{Scenario 1})}
\label{tabdelaisfiable}
\begin{tabular}{ | >{\centering} m{1.5cm} | m{2.05cm}|m{2.05cm} |m{2.05cm} |m{2.05cm}|} 
\hline
\rowcolor[gray]{0.75} 
{} &  \multicolumn{2}{c|}{\textbf{KmMC} } & \multicolumn{2}{c|}{\textbf{KmMR} }  \\
\cline{2-5}
\hline 
\rowcolor[gray]{0.85}
{DDoS attack rate}& 5 K8s Master Nodes & 7 K8s Master Nodes & 5 K8s Master Nodes & 7 K8s Master Nodes \\
 \hline
0 & 1701.91 & 2048.25 & 2746.45 & 3161.83 \\
\hline
2 & 2004.38 & 2132.93 & 2940.87 & 3179.45\\
\hline
4 & 2178.72 & 2471.39 & 3362.42 & 4521.79\\
\hline
4.5 & 2201.37 & 2501.73 & 3525.17 & 4632.38\\
\hline
5 & 2287.65 & 2623.87 & 3612.93 &  4729.98\\
\hline
5.5 & 2304.12 & 2702.99 & 3867.32 & 4970.93\\
\hline
6 & 2331.12 & 2732.25 & 4053.53 & 4970.93\\
\hline
\end{tabular}
\end {center}
\end{table*}
Considering \textit{Scenario 1}, we present in
Table \ref{tabdelaisfiable} the achieved consensus times versus DDoS attack rate of KmMC and KmMR, for a cluster of 5 and 7 master nodes respectively. For both platforms, consensus times increase slightly and proportionally to DDoS attack rates. Indeed, even with the additional Byzantine fault, $f$ and $f'$ respect the maximum number of tolerated faults\footnote{Notice that KmMC sees the DDoS attack as a crash event in this case.}. Hence, platforms' operation continue without significant degradation. However, KmMC realizes shorter consensus times than KmMR. This is expected, since the replication protocol Raft is designed with few consensus message exchanges between master nodes, compared to BFT-SMaRt. Finally, we conclude that it is recommended to select the KmMC platform if the risk of exceeding the maximum number of faults, dictated by Raft, is very low.

\begin{figure}  
  \centering
  \includegraphics[width=250pt]{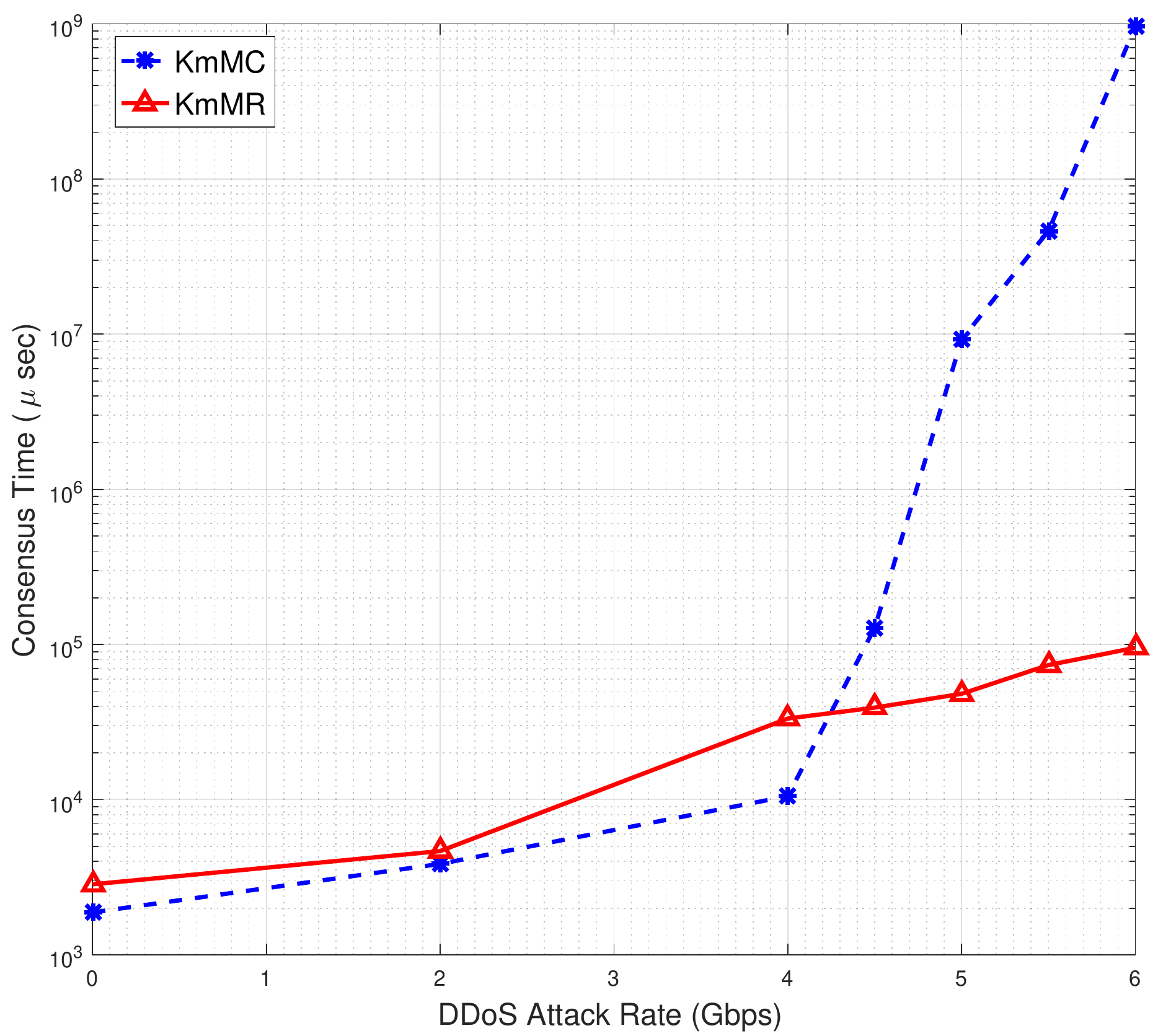}
  \caption{Consensus time versus DDOS attack rate (\textit{Scenario 2}, $n = 5$)}
  \label{5minstable}
\end{figure}

For \textit{Scenario 2}, we present in Fig. \ref{5minstable} the consensus time versus DDoS attack rate, for a cluster of 5 master nodes. The results show that the consensus time increases with DDoS attack rate. When the attack rate is below 4.25 Gbps, KmMC provides a slightly better performance than KmMR. Indeed, in this case, the DDoS victim resists to the attack thanks to its sufficient resources.
However, for an attack rate above 4.25 Gbps, KmMC deteriorates rapidly and significantly. This is mainly due to the vulnerability of Raft replication protocol in front of Byzantine faults. Indeed, the DDoS victim would behave improperly, e.g. not responding to other nodes in a timely manner. Thus, from this moment, Raft triggers changes in the cluster's leadership since it is no longer able to reach a consensus with its current \textit{leader}. This triggering considerably slows down consensus in the KmMC platform. 
Meanwhile, KmMR resists to all DDoS attacks, and is able to achieve consensus time 1000 times better than KmMC in average.
\begin{figure}  
  \centering
  \includegraphics[width=250pt]{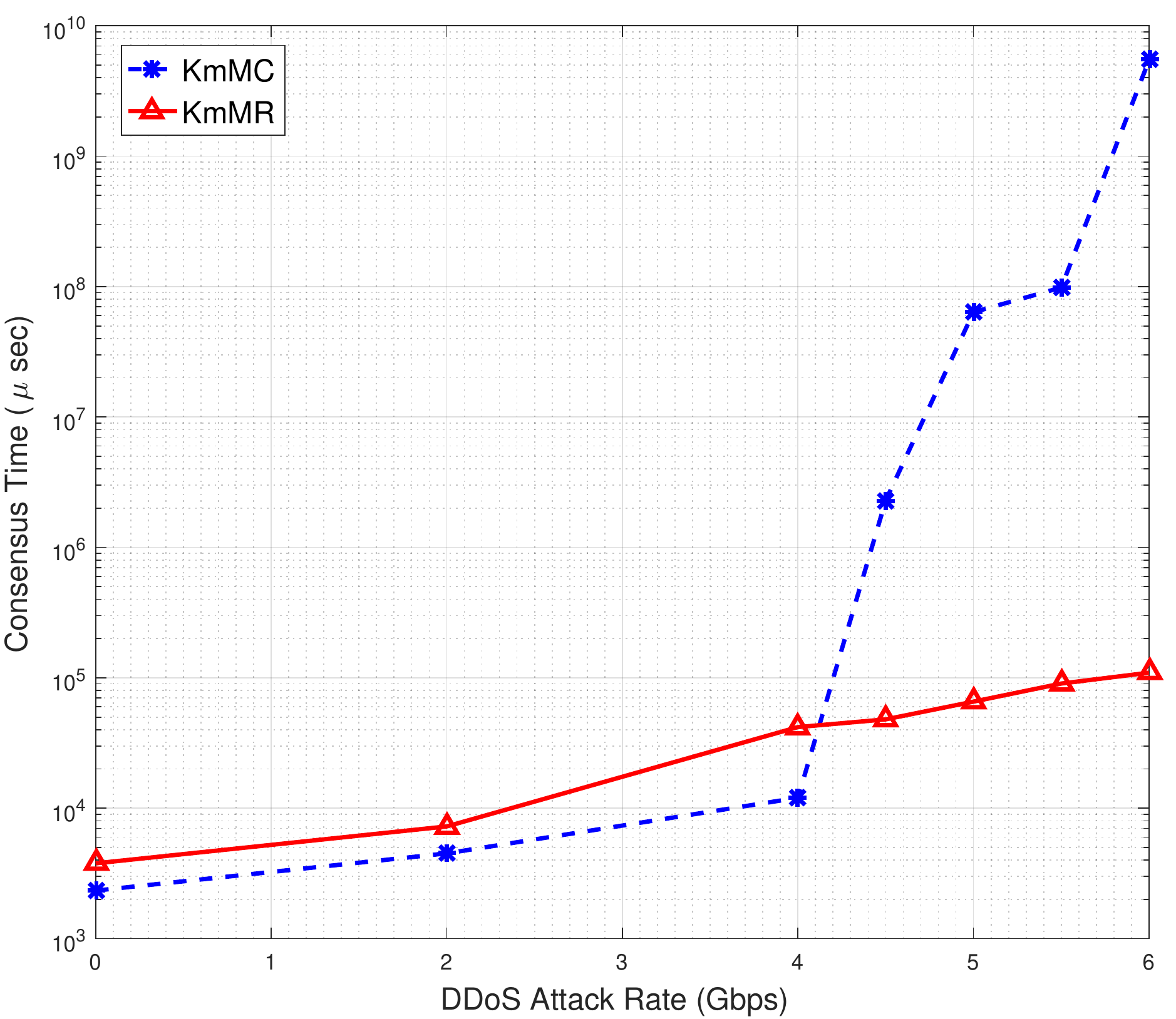}
  \caption{Consensus time versus DDOS attack rate (\textit{Scenario 2}, $n = 7$)}
  \label{7minstable}
\end{figure}

Fig. \ref{7minstable} illustrates the consensus time versus DDoS attack rate in the same environment as Fig. \ref{5minstable}, but for a cluster of 7 master nodes. The same behavior is exhibited for $n=5$ and $n=7$ master nodes. However, for $n=5$, consensus is established faster thanks to the smaller number of exchanged messages. 
As $n$ increases, KmMC becomes more susceptible to DDoS attacks. Indeed, the rapid degradation of KmMC's performance starts at attack rate 4.1 Gbps for $n=7$, compared to 4.25 Gbps for $n=5$.
Whereas, KmMR is able to establish consensus in a reasonable time, even for high attack rates.
\begin{figure}  
  \centering
  \includegraphics[width=250pt]{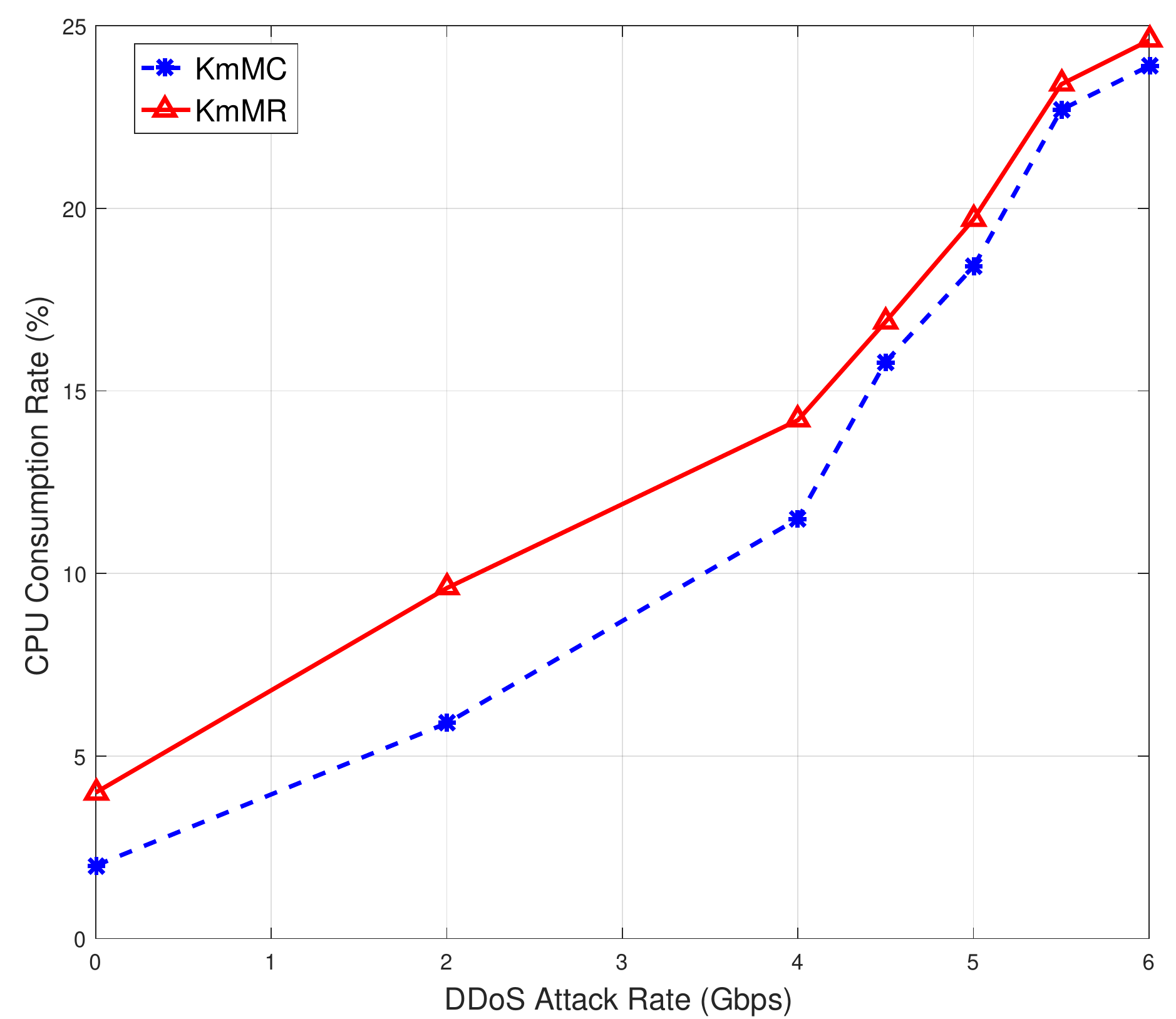}
  \caption{ CPU consumption rate versus DDOS attack rate (\textit{Scenario 2}, $n=7$)}
  \label{cpu}
\end{figure}
\begin{figure}  
  \centering
  \includegraphics[width=250pt]{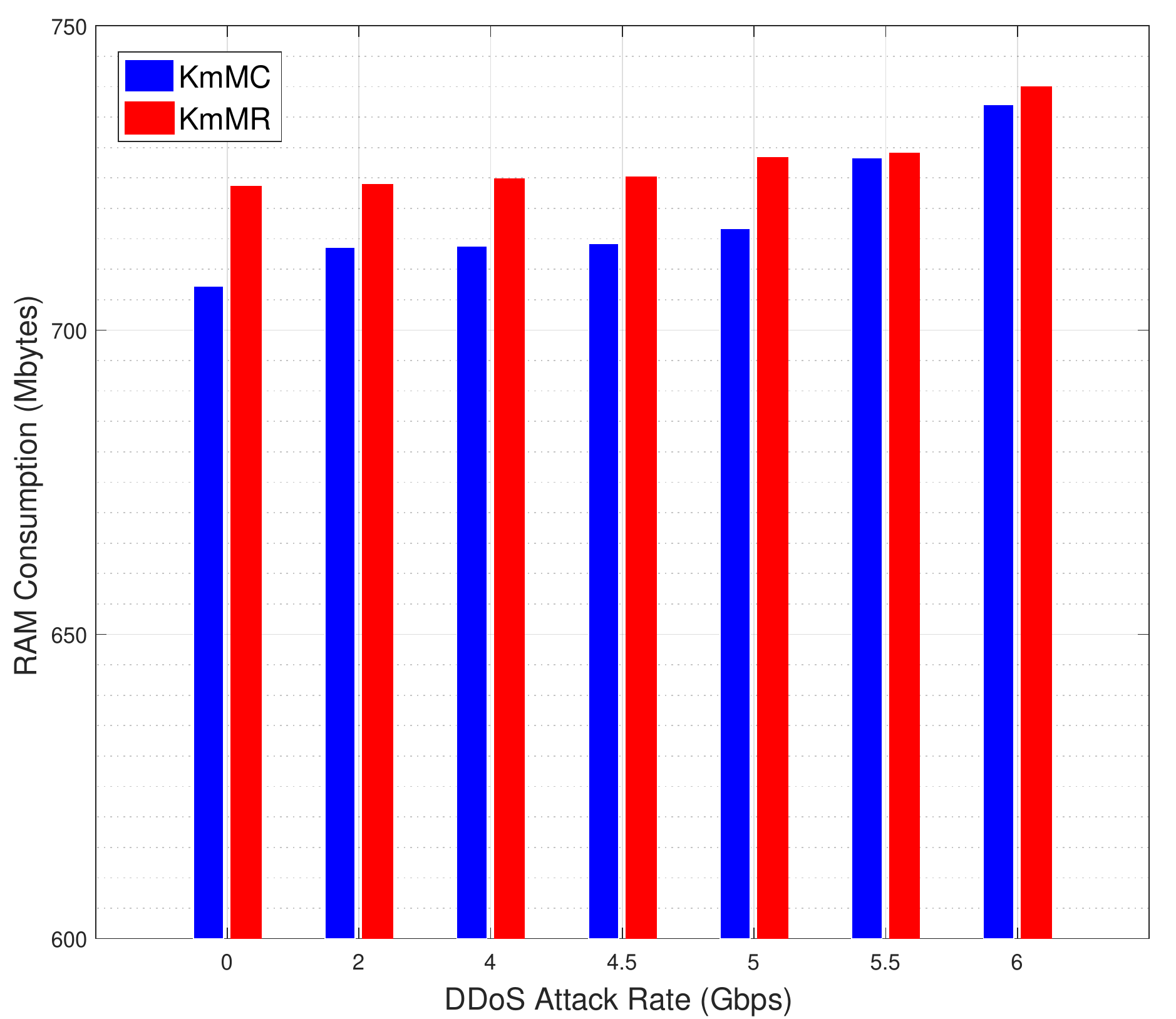}
  \caption{RAM consumption versus DDOS attack rate (\textit{Scenario 2}, $n=7$)}
  \label{ram}
\end{figure}
\begin{figure}  
  \centering
  \includegraphics[width=250pt]{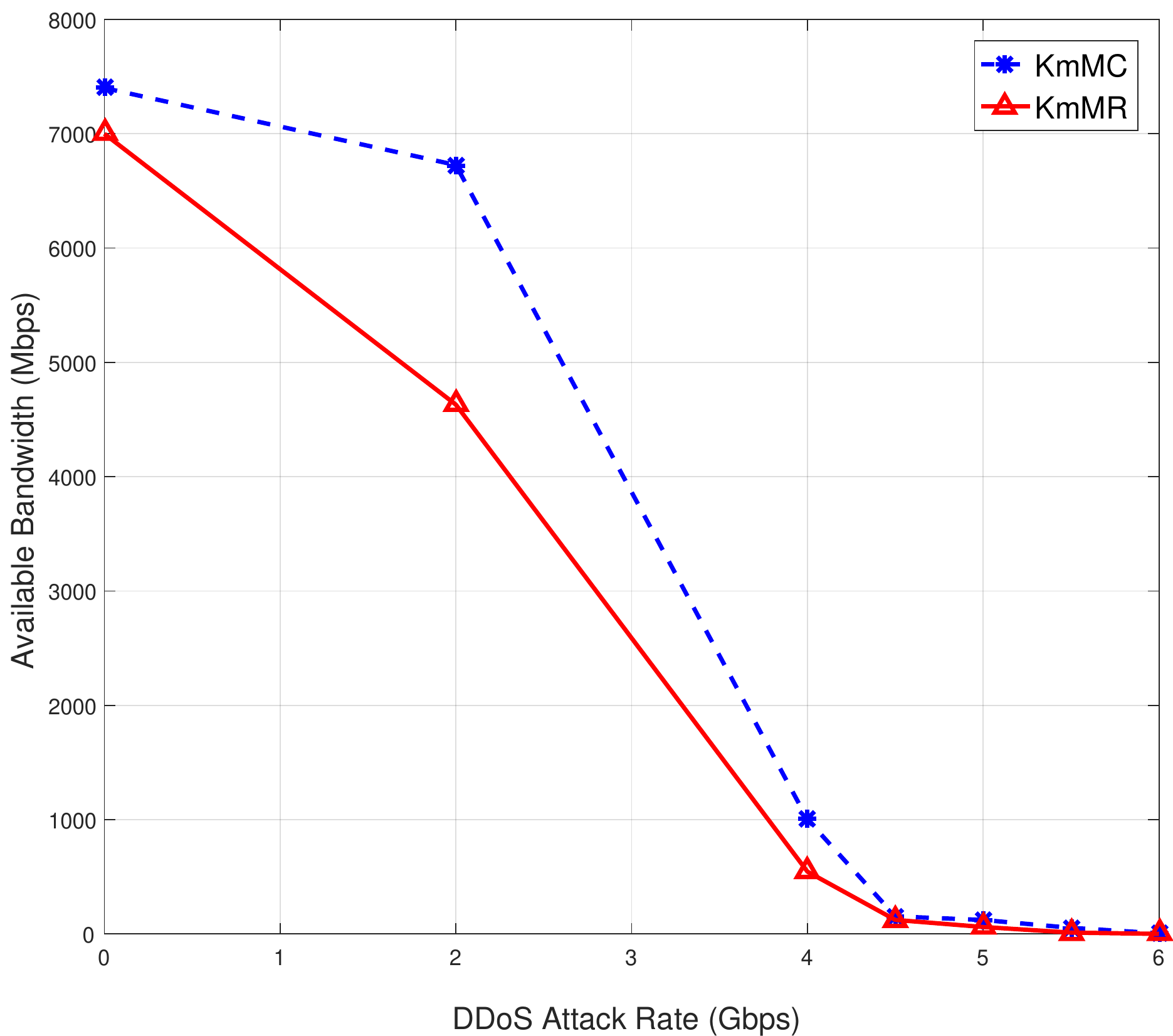}
  \caption{Available Bandwidth versus DDOS attack rate (\textit{Scenario 2}, $n=7$)}
  \label{bandepassante}
\end{figure}

Figs. \ref{cpu}-\ref{bandepassante} present the CPU, RAM and Bandwidth performances of the DDoS victim node, for \textit{Scenario 2} and $n=7$. When the DDoS attack rate is below 4.5 Gbps, KmMR uses as much or more resources than KmMC. This is expected since establishing a consensus in KmMR using BFT-SMaRt requires a larger number of messages exchange. However, for attack rates above 4.5 Gbps, KmMR and KmMC have almost the same level of resource utilization. Indeed, Raft starts to make changes in the cluster in order to regain its stability, resulting in higher resources consumption than usual. 

\textcolor{black}{
\subsection{Solution Limitations and Future Insights}
When researchers proposed PBFT-like protocols, such as BFT-Smart, their main concern was to enhance the performance of BFT in fault-free cases, while maintaining properties of Liveness and Safety, when faults occur. BFT-SMaRt aims to be robust in terms of high performance in fault-free executions, and correctness when faults happen. According to our experiments, it is clear that BFT-Smart is capable of surviving in a small and partially uncontrolled environment, where Byzantine faults may occur. However, it would reach rapidly its limits in a larger network, mainly due to its heavy communication and limited scalability. 
\\
To reinforce resistance to malicious Byzantine faults, the notion of robust BFT protocols has been introduced by Aardvark, i.e., maintaining a constant and stable performance in the presence of few Byzantine faults \cite{clement2009making}. Indeed, several improvements have been proposed, such as Aardvark \cite{clement2009making}, Spinning \cite{Vero2009}, and RBFT \cite{Aublin2013}, in order to efficiently handle some worst-case malicious Byzantine behaviors. For instance, Aardvark tolerates nodes/replicas to expect a minimum acceptable throughput from the leader \cite{clement2009making}, while Spinning changes the leader with every batch of requests \cite{Vero2009}. Finally, RBFT \cite{Aublin2013} proposed to maintain a constant performance during a fault event. It is demonstrated in \cite{Aublin2013} that Aardvark and Spinning performances are reduced by at least 78\% in presence of a fault, whereas RBFT degrades by only 3\%. This is due to RBFT's design, where $f+1$ protocol instances are ran, but only one executes the received request.
\\
Although interesting, the previous protocols would experience difficulties in managing inconsistency in large scale systems \cite{buchman2016tendermint}. Indeed, this type of management is relegated to client nodes, although the reason to use a consistent BFT protocol is precisely to avoid this responsibility to clients. One of the promising solutions is to concurrently run independent processes aiming at achieving higher throughputs \cite{Kotla2004}, which is the basic approach to implement scalable blockchain architectures. 
\\
Blockchain, by itself, is a BFT replicated state machine, where each state-update is a Turing machine with bounded resources. Unlike conventional BFT protocols where fault tolerance is realized among a small/medium group of nodes through rounds of message exchanges (votes and safety-proofs messages), blockchain achieves BFT among a very large number of participants, where at each time period, only a single message (Proof-of-Work -PoW- message) is broadcast by a participant. Adopting known BFT mechanisms into blockchain has led to the proposal of hybrid solutions, such as Byzcoin \cite{Kokoris2016}, Bitcoin-NG \cite{Eyal2016}, Casper \cite{buterin2017casper} and Solida \cite{abraham2016solida}. These approaches anchor off-chain BFT decisions inside a PoW chain or the other way around. For instance, Casper is a proof-of-stack (PoS)-based finality system, which overlays a PoW blockchain.
By design, Casper allows to provide Safety and plausible Liveness, as well as protect the system against \textit{long range revisions} and \textit{catastrophic crashes} faults \cite{buterin2017casper}.
Moreover, innovative solutions in the age of blockchains, such as Honeybadger \cite{Miller2016}, Algorand \cite{Yossi2017}, 
and LightChain \cite{hassan2019lightchain}, revisit the BFT setting with greater scalability and simplicity. Honeybadger is a demonstrative example of how BFT can build a blockchain cryptocurrency \cite{Miller2016}. It can reach consensus within 5 minutes using 104 nodes. By design, Honeybadger requires prior setting of a fixed number of consensus nodes, which may be problematic in terms of targeted attacks that may either compromise the nodes or exclude them from the system. In contrast, Algorand, a PoS approach, can achieve better performance without having to select a fixed set of nodes beforehand \cite{Yossi2017}. Also, it is robust against malicious attacks, even from a malicious leader, and scales better for a large number of clients. Finally, in order overcome the low communication and storage efficiency, inconsistency and scalability problems encountered in existing blockchains, the authors in \cite{hassan2019lightchain} proposed {LightChain}. The latter is a blockchain defined over a skip graph-based peer-to-peer distributed hash table overlay, which achieves consensus through Proof-of-Validation (PoV), i.e., a blockchain data is considered valid if its hash value is signed by a randomly selected number of validators \cite{hassan2019lightchain}. It has been proven that LightChain is a fair, consistent, and communication/storage efficient blockchain.   
\\
In spite of its limits, it is clear that the implementation of BFT-Smart into Kubernetes is the first step into providing robustness to this popular Docker containers orchestration platform. As future work, one could investigate the integration of more sophisticated protocols to Kubernetes, such as the aforementioned ones, and test their robustness to malicious Byzantine behaviours. Testing can be realized through the BFT-bench framework introduced in \cite{Gupta2016_2}.
}

\section{Conclusion}
With the increased importance of virtualization in cloud computing, Docker containerization is favored for its lightweight and efficient virtualization. This implies the emergence of new forms of architectures organizing cloud services in containers, ready to be instantiated in virtual and/or physical machines. Since the main objective is to guarantee service continuity, orchestrating these containers may seem challenging. Recently, Kubernetes has been adopted as the orchestration platform of Docker containers. Although efficient in managing containers, Kubernetes guarantees service continuity only in presence of non-Byzantine (crash) faults occuring within the system. In fact, the current replication protocol within Kubernetes ``Raft" cannot handle Byzantine faults. In this paper, we propose a new orchestration platform capable of overcoming this limitation in Kubernetes. The KmMR platform, based on Byzantine replication protocol BFT-SMaRt, is presented. We detailed our approach to integrate the BFT-SMaRt library (written in Java) into Docker and Kubernetes (written in Golang). Then, we implemented a Kubernetes multi-master platform in an OpenStack-based cloud environment. The system is evaluated for two different scenarios, where initially the maximum number of tolerated faults is either reached or not, and for two orchestration platforms, KmMC and KmMR. The results show that the conventional approach KmMC is efficient and robust in a non-Byzantine and controlled environment, i.e. number of maximum tolerated faults is not exceeded. However, in a Byzantine and not fully controlled environment, KmMR guarantees the continuity of services, while KmMC collapses in front of severe Byzantine faults. 
In a such environment, KmMR resources consumption is typically stable, compared to KmMC.  
In future works, we \textcolor{black}{will investigate the integration of more robust BFT protocols into Kubernetes, in order to ensure a better protection against malicious Byzantine faults.}


\section*{Acknowledgement}
This work was partially funded by NSERC-CRD program.
\bibliography{bib}
\end{document}